\documentclass{aa}

\usepackage{graphics}
\usepackage{psfig}

\begin{document}

\title {New Measurements of Orbital Period Change in Cygnus~X-3}
\author{ N. S. Singh\inst{1}, S. Naik\inst{2}, B. Paul\inst{2}, P. C. Agrawal\inst{2}, A. R. Rao\inst{2} and K. Y. Singh\inst{1}}

\offprints{S. Naik : sachi@tifr.res.in} 

\institute{Dept. of Physics, Manipur University, Canchipur, Imphal 795003, Manipur, India
           \and
Tata Institute of Fundamental Research, Homi Bhabha Road, Mumbai 400005, India}
\authorrunning{N. S. Singh et al.} 
\titlerunning{Orbital period change in Cygnus~X-3}

\abstract{
A nonlinear nature of the binary ephemeris of Cygnus~X-3 indicates either a 
change in the orbital period or an apsidal motion of the orbit. We have made 
extended observations of Cygnus~X-3 with the Pointed Proportional Counters 
(PPCs) of the Indian X-ray Astronomy Experiment (IXAE) during 1999 July 
3$-$13 and October 11$-$14. Using the data from these observations and the 
archival data from ROSAT, ASCA, BeppoSAX and RXTE, we have extended the data 
base for this source. Adding these new arrival time measurements to the 
published results, we make a comparison between the various possibilities, (a) 
orbital decay due to mass loss from the system, (b) mass transfer 
between the stars, and (c) apsidal motion of the orbit due to 
gravitational interaction between the two components. Orbital decay 
due to mass loss from the companion star seems to be the most probable 
scenario.
}
\maketitle

\section{Introduction}
The nature of the compact object in the bright X-ray binary Cygnus~X-3, 
is a subject of much debate. In spite of being one of the most
frequently observed X-ray sources, presence of an X-ray pulsar, a black hole,
or a low magnetic field neutron star has not yet been established. There is
also uncertainty about the mass and type of the companion star. An 
interesting way to probe this system is to 
investigate the arrival time history of the 4.8 hr orbital modulation 
in the X-ray light curve.

The unusual X-ray binary Cygnus~X-3 is located in the plane 
of our galaxy at a distance of $>$ 11.6 kpc (Dickey 1983). Because
of the high luminosity of the source in X-ray, infrared and radio bands, 
it has been observed on many occasions at these wavelengths. 
Strong optical extinction in the direction of the source prevents 
optical observations. The mass of the companion star has been determined 
by many different means  and it is found to be in the range of a 
fraction of solar mass to a few solar mass (Van den Heuvel \& De 
Loore 1973; Tavani, Ruderman, \& Shaham 1989; van Kerkwijk et al. 
1992). Parsignault et al. (1976) discovered a periodicity of 4.8 hr 
in the X-ray flux with nearly sinusoidal variation, which is believed 
to be due to the orbital motion. The variation in the sinusoidal shape of 
the light curve from cycle to cycle is reported by van der Klis and 
Bonnet-Bidaud (1981). They showed that the shape of the average light 
curve formed from successive cycles is quasi-sinusoidal with a slow rise 
and fast fall. Several models have been proposed to explain the deep 
and near-sinusoidal modulation of the X-ray intensity of Cygnus~X-3 with 
the orbital period (Ghosh et al. 1981). Material in the form of a large 
shell around the binary system can produce the X-ray modulation by 
scattering, absorbing, and re-emitting the X-rays from the compact object, 
if the X-ray emission is only from the non-shadowed part of the shell.
White \& Holt (1982) proposed the accretion disk corona (ADC) model to 
explain the modulation in the X-ray light curve of the source. According 
to this model, the X-ray source is covered by an optically thick corona 
with a radius of about 10$^9$ cm, an optical depth of about 10 and a 
temperature of about 2 keV. Another model which tries to explain the orbital 
modulation is stellar wind model (Willingale, King, \& Pounds 1985; Kitamoto 
et al. 1987). This model assumes a strong and highly ionized stellar wind 
from the companion star with an optical depth of about 1 and a radius of 
about 10$^{11}$ cm and the temperature of the wind is model dependent.

Evolution of the orbit of Cygnus~X-3 is studied by measuring the arrival 
times of the minima in each orbital motion. The ephemeris of 4.8 hr 
modulation in the X-ray light curve of the source has been  
studied by many authors (Leach et al. 1975; Mason \& Sanford 1979; 
Parsignault et al. 1976; Lamb, Dower \& Fickle 1979; Elsner et al. 
1980; van der Klis \& Bonnet-Bidaud 1981, 1989; Kitamoto et al. 1987). 
The time derivative of the orbital modulation period ($\dot{P}$) and
second derivative ($\ddot{P}$) have been measured as $\sim$ 10$^{-9}$
s s$^{-1}$ and $\sim$ -10$^{-11}$ yr (van der Klis 
\& Bonnet-Bidaud 1981, 1989; Kitamoto et al. 1987). 
The unusually large value of $\dot{P}$/P (= 2.2 $\times$
10$^{-6}$ yr$^{-1}$ ), which if linked to the binary evolution of the system,
is similar to several short-period bright LMXBs such as
X1822$-$371 with $\dot{P}$/P $\sim$ 3.4 $\times$ 10$^{-7}$ yr$^{-1}$ and
orbital period of 5.57 hr (Hellier et al. 1990; Parmar et al. 2000), 
EX00748$-$767 with $\dot{P}$/P $\sim$ 2 $\times$ 10$^{-7}$ yr$^{-1}$ 
and orbital period of 3.82 hr (Parmar et al. 1991).
This value of $\dot{P}$ can be interpreted as due to 
the mass transfer from the optical companion to the compact object. 
Using the measured period derivative of Cygnus X$-$3, Kitamoto 
et al. (1987) determined the rate of mass loss from the binary companion as 
about 10$^{-6}$ M$_{\odot}$ yr$^{-1}$. 
Tavani et al. (1989) tried to explain the
observed high value of $\dot{P}$/P in Cygnus~X-3 and the required mass
transfer rate close to $\sim$ 1.58 $\times$ 10$^{-8}$ 
M$_{\odot}$ yr$^{-1}$. They claimed that the
companion is a degenerate star with solar composition and in the mass
range of 0.01 $\leq$ $m$ $\leq$ 0.03 M$_{\odot}$ which under-fills its
Roche lobe. The X-ray illumination from the primary compact object
produces the evaporative wind from such a low-mass degenerate companion.
However, from infrared observations, van Kerkwijk et al. (1992) discovered
that the binary companion of Cygnus~X-3 has the spectrum of a Wolf-Rayet
star and predicted the mass of the companion as $\sim$ 10 M$_{\odot}$
which suggests that Cygnus~X-3 is a high mass X-ray binary. Recent 
observations by Fender et al. (1999) confirm the nature of
the companion as an early-type WN Wolf-Rayet star.

We have measured the available arrival time of minimum value of the light 
curve from two observation of Cygnus~X-3 with PPCs of IXAE in 1999 and 
several archival data. By combing these new measurements with the previously 
published results, we are able to determine the rate of change of orbital
period with greater accuracy.

\section{Observations and Archival Data}  
\subsection{PPC Observations}
The X-ray observations of Cygnus~X-3 were carried out twice on 
1999 July 3$-$13 and 1999 October 11$-$14 with 1 s time-integration mode
with the PPCs of the IXAE on board the Indian satellite IRS-P3. The IXAE 
includes three co-aligned and identical, multi-wire, multi-layer proportional 
counters with a total effective area of 1200 cm$^{2}$ covering 2 to 18 keV 
energy range with an average detection efficiency of about 60\% in 3$-$7 keV 
energy range. Each PPC is equipped with a honeycomb type collimator having a 
field of view of 2$^{\circ}$.3 $\times$ 2$^{\circ}$.3 and has an overall energy
resolution of about 22\% at 6 keV (Agrawal et al. 1998; Rao et al. 1998). 
The IRS-P3 satellite is in a circular orbit at an altitude of 830 km and 
inclination of 98$^{\circ}$. Pointing towards any particular source is done 
by using a star tracker with an accuracy of \(\leq\) 0$^{\circ}$.1. The useful 
observation time is limited to the latitude range typically from 30$^{\circ}$S 
to  50$^{\circ}$N to avoid the rise in charged particle background at high 
latitudes. The South Atlantic Anomaly (SAA) region restricts the observation 
to 5 of the 14 orbits per day. As the detectors are co-aligned, simultaneous 
background observation is not possible. The background counts are measured 
after the observation of a source by pointing the PPCs to a source-free region 
in the sky, close to the target source. The total useful exposure time of 
the X-ray source during the two observations is 31245 s. The PPC light curves 
did not show strong intensity variations on the time scale of a few seconds to 
a few hundred seconds.

\subsection{Archival data}

\setcounter{figure}{0}
\begin{figure*}
\vskip 11.0cm
\includegraphics{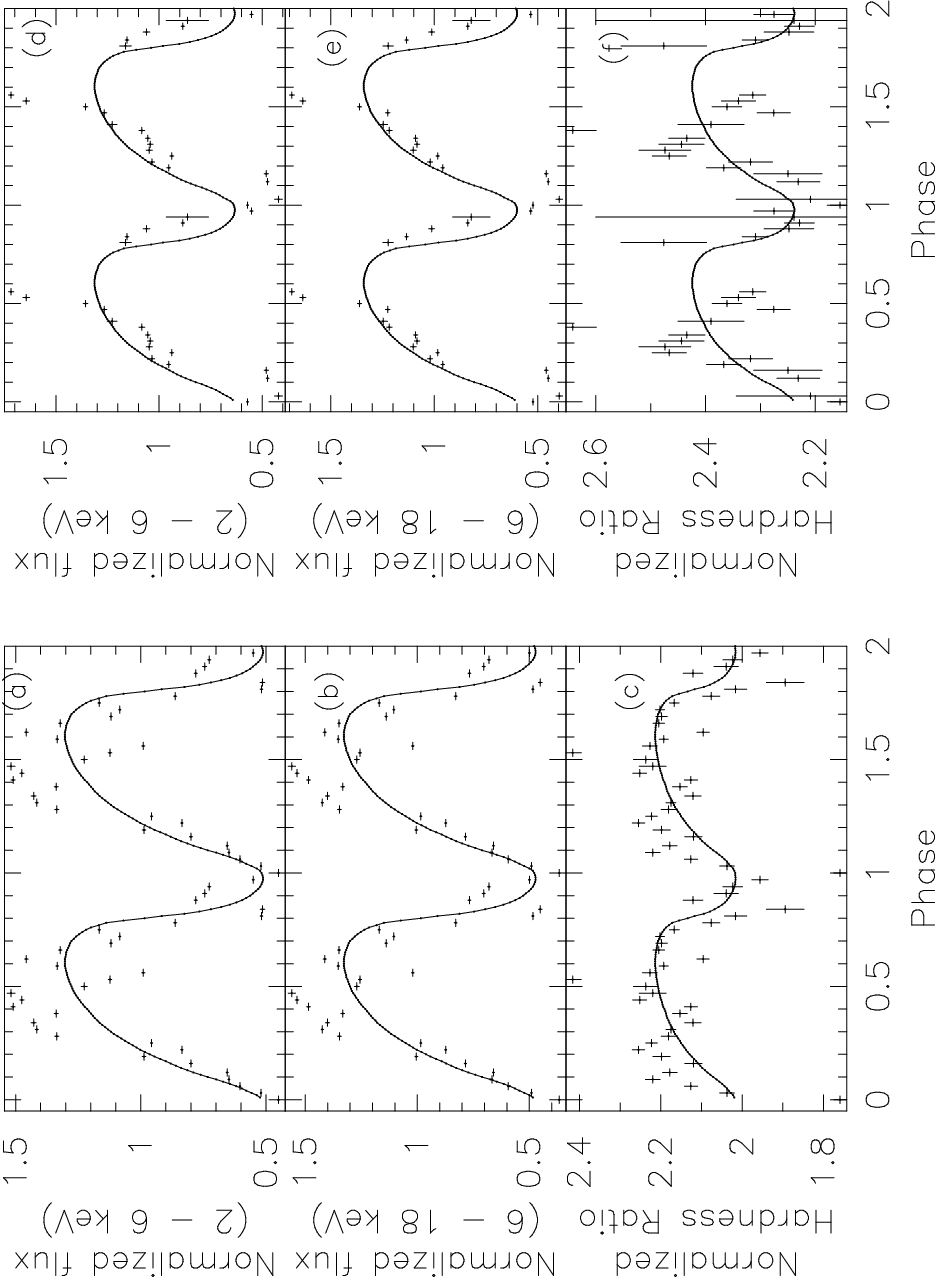}
\caption{Folded light curves in 2$-$18 keV and 6$-$18 keV energy ranges
are shown in the panels (a) and (b) (for 1999, July 3$-$13 PPC observations) and
panels (d) and (e) (for 1999, October 11$-$14 PPC observations) on a normalized
scale with the template data (described in text) which is indicated by lines.
The panels (c) and (f) show the folded hardness ratios (ratio of the count 
rate in 6$-$18 keV energy range to the count rate in 2$-$6 keV energy range 
for 1999, July and October observations respectively.}
\end{figure*}

We have also analyzed  archival data from several satellites. From ROSAT 
mission, we have analyzed HRI data with two different observation IDs.       
The durations of these two observations are inter-wined, but the overall 
count rate is different because in one of the observations, the source was
near the edge of the field of view where effective area is small. RXTE/PCA  
observations of the source in 2$-$60 keV energy range, combined data from GIS 
detectors (GIS2 and GIS3) and SIS detectors (SISO and SIS1) of the ASCA
satellite and data from MECS detectors of BeppoSAX observations were used 
to determine more arrival time information of Cygnus~X$-$3. We have analyzed 
the RXTE/ASM dwell data in the energy range of 2$-$12 keV for the entire 
period of observation starting from 1996 January to 2001 August 30.
The details of the observations of the source with different satellites
are given in Table 1 along with the useful exposure period.

\setcounter{table}{1}
\begin{table}[h]
\caption{Archival data used for the determination of the arrival time of
the minima of the binary modulation}
\begin{center}
\begin{tabular}{lllll}
\hline
\hline
Satellite  &Date of observation  &Exposure  &Detector\\
	   &        &(in seconds)          & \\
\hline
\hline
ASCA       &1994 29/5 $-$ 30/5   &21468 &GIS \\
ASCA       &1994 29/5 $-$ 30/5   &17527 &SIS \\
ROSAT      &1995 20/4 $-$ 22/4   &72709 &HRI\\
ROSAT      &1995 21/4 $-$ 23/4   &51670 &HRI\\
BeppoSAX   &1996 22/9 $-$ 24/9   &51626 &MECS \\
BeppoSAX   &1996 10/10           &26167 &MECS \\
BeppoSAX   &1996 29/10           &13664 &MECS \\
BeppoSAX   &1997 19/9 $-$ 20/9   &25802 &MECS \\
RXTE       &1996 24/8 $-$ 30/8   &86450 &PCA \\
RXTE       &1997 16/2 $-$ 22/2   &59465 &PCA\\
RXTE       &1997 05/6 $-$ 26/9   &64693 &PCA\\
RXTE       &1998 16/5 $-$ 21/5   &34723 &PCA\\
RXTE       &1996 04/1 $-$ 	 &\\
	   &2001 30/8  &  &ASM \\
\hline
\hline
\end{tabular}
\end{center}
\end{table}

\section{Data Analysis and Results}
\subsection{Arrival time determination}

The data used for the analysis were corrected to the solar system 
barycenter. To avoid the cycle to cycle variation in the light curve, 
we have created the time averaged orbital modulation profile by folding 
the source light curves with a period of 17253 seconds (Elsner et al. 1980). 
In all cases, the shape of the folded light curves is somewhat similar to 
the shape of the template function which is the long term average shape of the 
light curve obtained with the Copernicus satellite (Mason et al. 1976).

To determine the arrival times of the minimum of the folded light curves, 
we have used the normalized template data of X-ray ephemeris of Cygnus~X-3 
(van der Klis et al. 1989). The folded data were cross-correlated with the 
standard template function as described by van der Klis et al. (1989) to get 
the phase difference in the minima of the template and the data which 
give the measure of the arrival time of the minima. In addition to 
the statistical variations, Cygnus~X-3 often shows some short-term intensity
variations superposed on the orbital modulation.  We have multiplied the
statistical error estimates of each folded light curve by some constant
factors to make them fit to the template with a reduced $\chi^2$ of 1.0.
The error in the arrival times were estimated from the uncertainties in 
determining the center of the Gaussian function of the cross-correlated 
data. The short-term occasional intensity variations also cause some random
fluctuations in the arrival time determinations. Following the earlier
works (Kitamoto et al. 1995), we have quadratically added 0.002 d to the 
arrival uncertainties to make them compatible with the previous reported
measurements.

\setcounter{figure}{1}
\begin{figure}
\vskip 7.5cm
\includegraphics{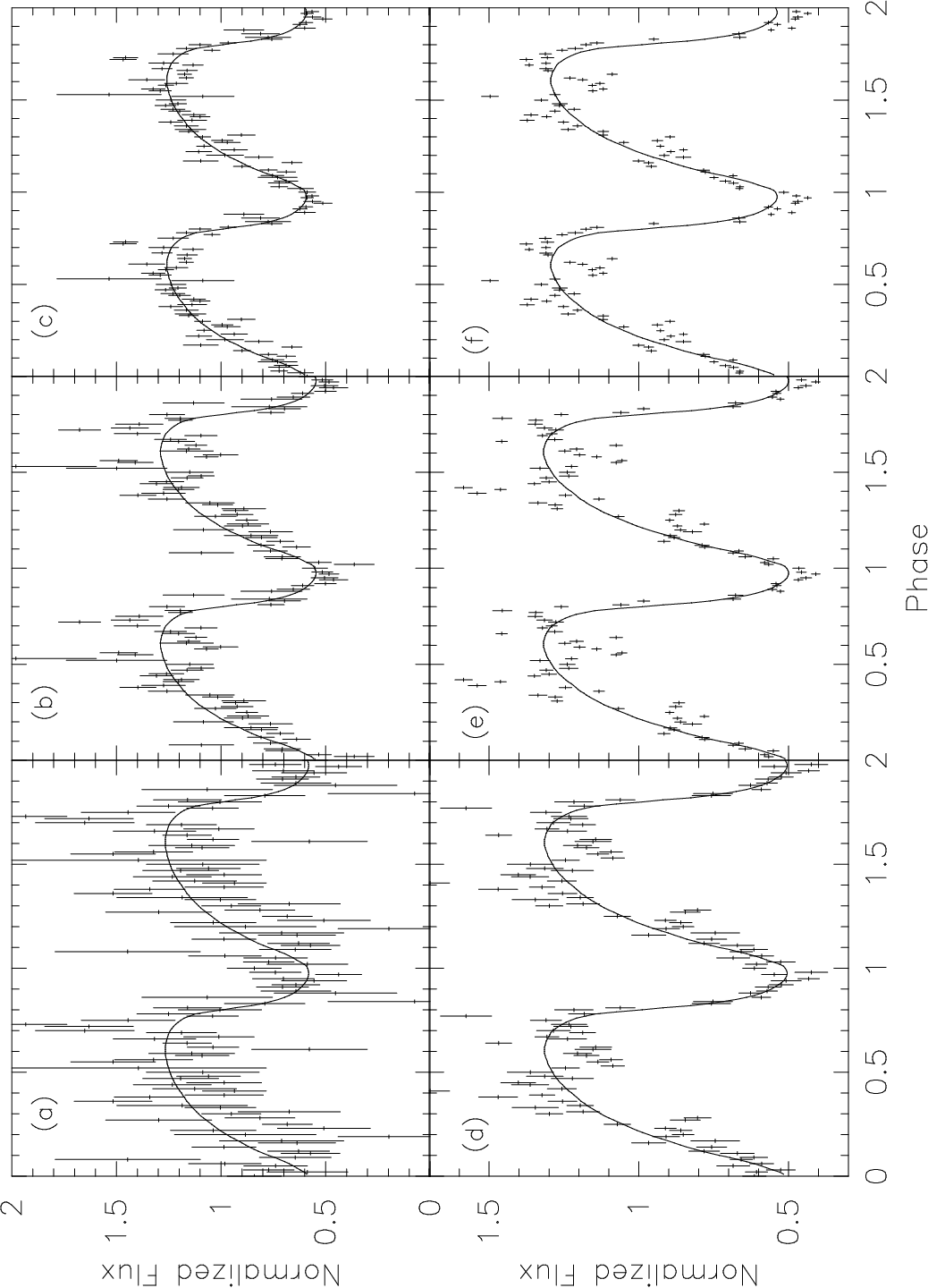}
\caption{The figure shows the normalized folded light curves of the source
obtained from the RXTE/ASM data for different energy ranges and at different
time. Upper panels (panels a, b, and c) show the folded light curves
in the energy ranges 1.3$-$3.0 keV, 3.0$-$5.0 keV, and 5.0-12.1 keV for the
data in the time range JD 2450086.8222 $-$ 2450186.4878 whereas the lower
panels (panels d, e, and f) show the folded light curves in the energy ranges
1.3$-$3.0 keV, 3.0$-$5.0 keV, and 5.0-12.1 keV for the data in the time range
JD 2450887.2862 $-$ JD 2450986.1345 respectively.}
\end{figure}

This analysis yielded two new arrival points from PPC observations 
(Singh et al. 2001), two 
points from ASCA, two points from ROSAT, five points from RXTE/PCA, and 
four points from BeppoSAX observations. We divided the RXTE/ASM light 
curve into 19 segments of durations of 100 days as given in Table 2, 
and determined the arrival times from each data segment. The information 
regarding the new arrival times are given in Table 1 along with the 
observation periods of the source with different instruments on different 
satellites and the orbit number. In accordance with the earlier works
(Leach et al. 1975), we have assigned the orbit number of the source as 0
at JD 2440949.9201. The orbit numbers for later observations were calculated 
by using 4.8 hour orbital period of the source. The significant increase in 
the number of the new arrival points to the old data improved the numerical 
values of the period derivative and the double derivative. 

\setcounter{figure}{2}
\begin{figure}
\vskip 11.5cm
\includegraphics{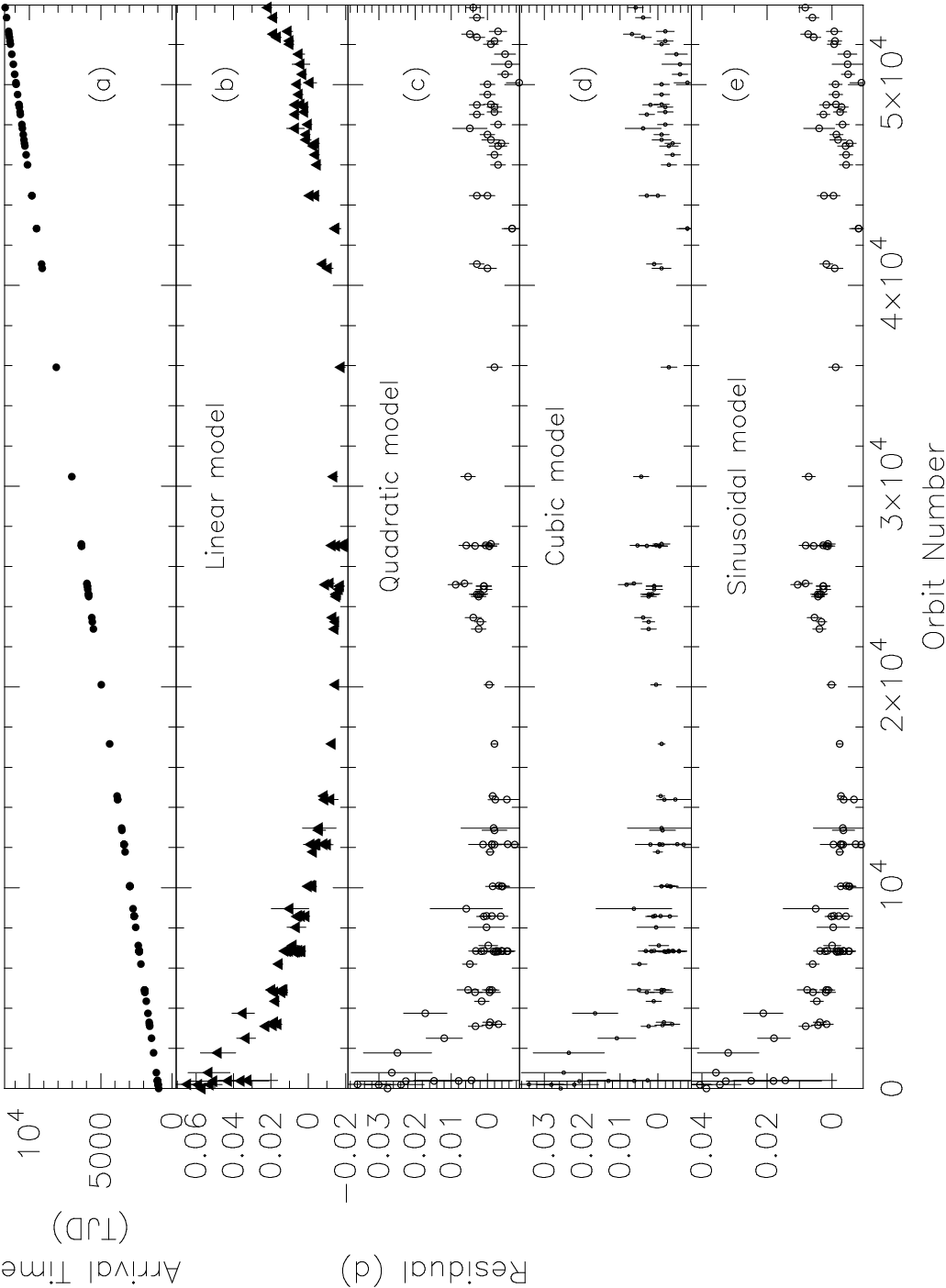}
\caption{The figure shows the arrival time history for the source Cygnus~X-3
(top panel). The data points beyond orbit number 42830 are obtained from
the present work. Arrival time residuals of the orbital modulation of the 
source with respect to the best-fit linear, quadratic, cubic and a linear and
sinusoidal models to the data are shown in different panels.}
\end{figure}

The orbital modulation of the light curves obtained from the observations 
of the source with different instruments are shown in different figures 
along with the normalized template. Panel (a) and (d) of Figure 1 show the 
folded light curves in 2$-$18 keV energy range along with the normalized 
template for 1999 July 3$-$13 and 1999 October 11$-$14 PPC observations 
of the source respectively. In Figure 2, we have shown the orbital 
modulation of the light curves along with the normalized template for two 
data segments of the RXTE/ASM for different energy bands. From these 
figures, it is observed that the shape of the orbital modulated light 
curves of Cygnus~X-3, obtained from various observations, is similar to 
the shape of the normalized template.

\setcounter{figure}{3}
\begin{figure}
\vskip 10.7cm
\includegraphics{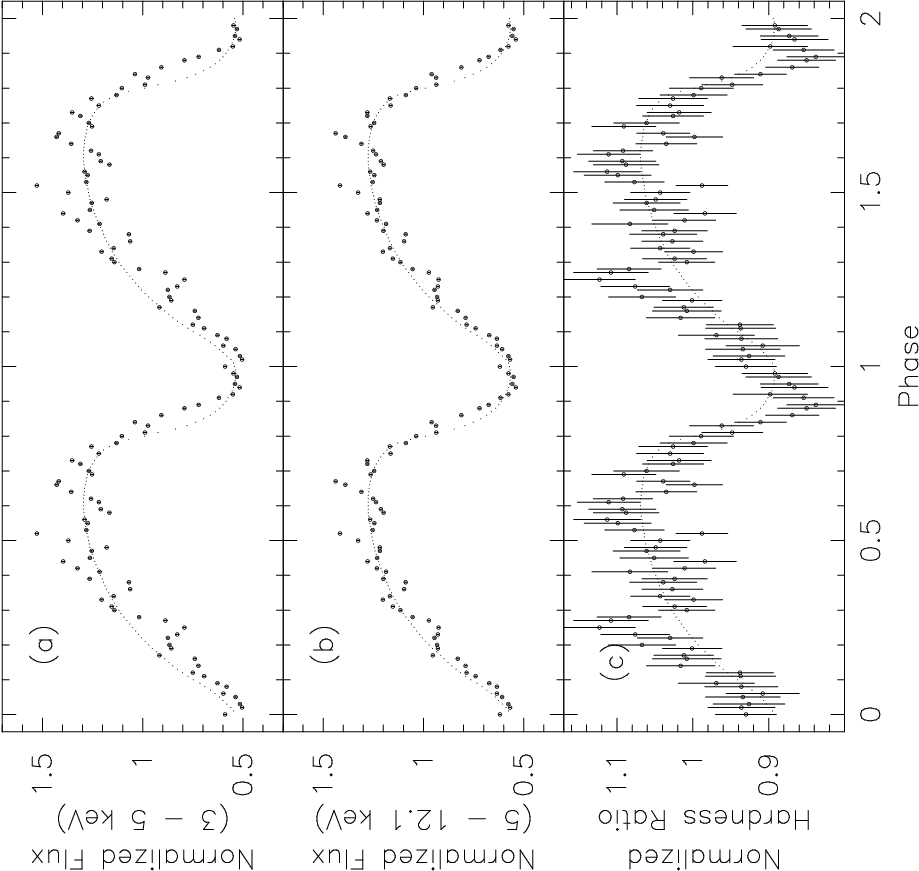}
\caption{The folded light curves for Cygnus~X-3 with RXTE/ASM in the energy
range 3$-$5 keV and 5$-$12.1 keV energy ranges along with the normalized 
template data are shown in panels (a) and (b). The panel (c) shows the folded
hardness ratio of the source plotted on a normalized scale with the template
data.}
\end{figure}

\begin{table*}
\centering
\caption{The summary of the observations used to determine the
ephemeris of Cygnus~X-3 (New record)}
\begin{tabular}{lllll}
\hline
\hline
Observation   &Satellite    & Number of   &Arrival time       &Error in  \\
Period        &(detector)   & Orbits      &(2440000 JD+)       &arrival time \\
(MJD)         &             & (p=17253 s) & (day)             & (day)      \\
\hline
\hline
49501.91$-$49502.84  &ASCA(GIS)      &42832     & 9502.8736       &0.0021\\
49501.91$-$49502.84  &ASCA(SIS)      &42832     & 9502.8736       &0.0028\\
49827.99$-$49829.44  &ROSAT(PSPC)    &44466     & 9829.1748       &0.0020\\
49828.59$-$49830.56  &ROSAT(PSPC)    &44470     & 9829.9766       &0.0021\\
50086.32$-$50185.98  &RXTE(ASM)      &46006     & 10136.6934      &0.0020\\
50186.06$-$50285.99  &RXTE(ASM)      &46506     & 10236.5381      &0.0020\\
50319.46$-$50325.83  &RXTE(PCA)      &46940     & 10323.2031      &0.0024\\
50348.92$-$50350.18  &BeppoSAX(MECS1) &47074    & 10349.9609      &0.0020\\
50385.00$-$50385.52  &BeppoSAX(MECS)  &47253    & 10385.7090      &0.0025\\
50386.31$-$50485.96  &RXTE(ASM)      &47508     & 10436.6299      &0.0020\\
50495.02$-$50501.12  &RXTE(PCA)      &47818     & 10498.5391      &0.0047\\
50486.02$-$50585.87  &RXTE(ASM)      &48008     & 10536.4727      &0.0020\\
50586.16$-$50685.95  &RXTE(ASM)      &48509     & 10636.5234      &0.0020\\
50604.75$-$50717.50  &RXTE(PCA)      &48635     & 10661.6787      &0.0021\\
50710.65$-$50711.26  &BeppoSAX(MECS2) &48884    & 10711.4014      &0.0020\\
50731.36$-$50731.90  &BeppoSAX(MECS2) &48988    & 10732.1738      &0.0021\\
50686.01$-$50785.73  &RXTE(ASM)      &49009     & 10736.3643      &0.0020\\
50786.51$-$50883.86  &RXTE(ASM)      &49506     & 10835.6104      &0.0021\\
50886.79$-$50985.63  &RXTE(ASM)      &50012     & 10936.6533      &0.0020 \\
50949.63$-$50954.98  &RXTE(PCA)      &50093     & 10952.8203      &0.0035\\
50986.41$-$51085.89  &RXTE(ASM)      &50513     & 11036.6934      &0.0020\\
51086.09$-$51185.95  &RXTE(ASM)      &51013     & 11136.5381      &0.0047 \\
51186.22$-$51285.96  &RXTE(ASM)      &51514     & 11236.5830      &0.0029\\
51286.03$-$51385.62  &RXTE(ASM)      &52013     & 11336.2324      &0.0021\\
51361.49$-$51371.77  &IXAE(PPC)      &52168     & 11367.1836      &0.0021\\
51404.44$-$51404.64  &RXTE(PCA)      &52358     & 11405.1309      &0.0022\\
51386.16$-$51485.96  &RXTE(ASM)      &52515     & 11436.4844      &0.0023\\
51461.72$-$51464.79  &IXAE(PPC)      &52651     & 11463.6338      &0.0024\\
51549.73$-$51649.73  &RXTE(ASM)      &53334     & 11600.028320    &0.0020\\ 
51649.90$-$51749.80  &RXTE(ASM)      &53836     & 11700.274410    &0.0020\\
51750.06$-$51850.02  &RXTE(ASM)      &54338     & 11800.516600    &0.0020\\
51850.08$-$51948.81  &RXTE(ASM)      &54836     & 11899.958010    &0.0020\\ 
51951.46$-$52051.45  &RXTE(ASM)      &55347     & 12002.000980    &0.0020\\
52051.58$-$52151.53  &RXTE(ASM)      &55848     & 12102.048830    &0.0020\\     
\hline
\end{tabular}
\end{table*}

\subsection{Orbital evolution of the binary system}

Figure 3 (top panel) reports the new measurements of the 
arrival time of the minima in the orbital modulation of the X-ray light 
curves of Cygnus~X-3 together with those of all the previously published 
values (Kitamoto et al. 1987; van der Klis \& Bonnet-Bidaud 1989).
We have tried to fit the arrival time data with 
different polynomial functions. The residuals of the various polynomial 
fits performed on these data are also shown in other panels of the figure. 
The second panel of Figure 3 shows the residual of the linear fit to the 
arrival data whereas the third, fourth, and fifth panels of Figure 3 show 
the residuals of the quadratic, cubic, and sinusoidal fit to 
the arrival time data respectively. 

The results of the various polynomial fits performed on the complete set
of arrival time data are listed in Table 3. The arrival times of the 
orbital modulations clearly deviate from a linear relationship with the 
orbit numbers. Addition of a quadratic term fits the data much better 
with a reduced $\chi^2$ of 3.08 (113 dof). From the quadratic fit, the 
value of the period derivative is derived to be $\dot{P}$ = 5.76333 
$\times$ 10$^{-10}$ i.e, an evolution time scale of $\dot{P}$/P=1.05419 
$\times$ 10$^{-6}$ yr$^{-1}$. These values are similar to what was 
obtained with the data prior to 1993 (Kitamoto et al. 1995). A cubic 
model for the arrival time fits the data slightly better with reduced 
$\chi^2$ of 2.96 (112 dof). However, the rate of change of $\dot{P}$ is 
found to be much smaller compared to that obtained by Kitamoto et al. 
(1995), $\ddot{P}$ =-1.34008 $\times$ 10$^{-11}$ yr$^{-1}$. The curved 
feature of the residual of the linear fit to the data prompted us to fit 
with a model consisting of linear and sinusoidal terms as model components. 
Although the reduced $\chi^2$ for this fitting is comparable to the 
reduced $\chi^2$ for cubic model, the amplitude of sinusoidal component 
is unphysical because, according to the method described by Batten (1973), 
it requires the eccentricity of the binary system to be $\gg$ 1. In the linear 
plus sinusoidal model, a reduced $\chi^2$ of 3.5 (dof 112) was obtained by 
restricting the amplitude of the sinusoidal component within a physical limit 
so that $e \le 1$. However, if a precessing eccentric orbit is the reason 
for the non-linear nature of the arrival times, a change in the shape 
of the orbital modulation of the light curve is expected, which has 
not been observed in the last $\sim$ 30 years. Based on this argument, 
apsidal motion seems not to be an appropriate description of the 
arrival time data.

\begin{table*}
\caption{Statistics of different fitting models to the data of arrival time }
\centering
\begin{tabular}{ll}
\hline
\hline
Linear ephemeris: $\chi^2$ = 4099 for 114 dof  & \\
$T_n$ = $T_0$ + P$n$                          & \\
$T_0$ = 2440949.863 $\pm$ 0.002 $JD$                    &  \\
$P$   = 0.19968775 $\pm$ 0.00000007 d       &Cov($T_0$, P) = -2.3 $\times$ 10$^{12}$\\
\hline
Parabolic ephemeris: $\chi^2$ = 348.3 for 113 dof     &\\
$T_n$ = $T_0$ + $P_0n + cn^2$  &Cov($T_0, P_0) = -2.3 \times 10^{-11} d^2$\\
where c =$P_0 \dot{P}$/2   &Cov($T_0, c) = 3.5 \times 10^{-16} d^2$ \\
$T_0$ = 2440949.892 $\pm$ 0.001 $JD$ &Cov($P_0, c) = -3.8 \times 10^{-20} d^2$\\
$P_0$ = 0.19968443 $\pm$ 0.00000009 d     &$\dot{P}$ = (5.76 $\pm$ 0.24) $\times 10^{-10}$    \\
$c$ = (5.75 $\pm$ 0.16) $\times$ 10$^{-11}$  d &$\dot{P}/P_0$ = (1.05 $\pm$ 0.04) $\times 10^{-6} yr^{-1}$\\
\hline
Cubic ephemeris: $\chi^2$ = 331 for 112 dof  &\\
$T_n$ = $T_0$ + $P_0n + c_0n^2 + dn^3$  &Cov($T_0, P_0) = (-9.1) \times 10^{-11} d^2$ \\
where c$_0$ =$P_0 \dot{P}$/2  and  &Cov($T_0, c_0) = 3.1 \times 10^{-15} d^2$ \\ 
$d \simeq P^{2}_0 \ddot{P}/6$   &Cov($T_0, d) = (-3.2) \times 10^{-20} d^2$  \\
$T_0 = 2440949.8944 \pm 0.0015$ JD    &Cov($P_0, c_0) = (-5.5) \times 10^{-19} d^2$\\ 
$P_0$ = 0.199684009 $\pm$ 0.0000003 d        &Cov($P_0, d) = 5.9 \times 10^{-24} d^2$\\
$c_0 = (7.7 \pm 1.1) \times 10^{-11}  d$  &Cov($c_0, d) = (-2.4) \times 10^{-28} d^2$\\
$d = (-2.4 \pm 1.3) \times 10^{-16}  d$  &$\dot{P} = (7.7 \pm 1.5) \times 10^{-10}$\\
       &$\ddot{P} = (-1.3 \pm 1.6) \times 10^{-11} yr^{-1}$ \\
\hline
Linear $+$ Sinusoidal ephemeris:  $\chi^2$ =  391.1 for 112 dof \\
$T_n$ = $T_0$ + $P_0n + c_0~sin(2\pi(n-\phi)/P_{aps})$\\
$T_0$=2440949.92 $\pm$ 0.03 JD\\
$P_0=0.199687 \pm 9 \times 10^{-07} d$\\
$c_0=0.059 \pm 0.013 d$\\
$\phi=334500 \pm 6000 $  no. of orbit\\
$P_{aps}$=75.4 $\pm$ 0.2 yr \\
\hline
\hline
\end{tabular}
\end{table*}       

\section{Spectral variations with orbital phases in Cygnus~X-3}

To investigate the spectral variations of the source during different
orbital phases, we have analyzed the hardness ratios from the PPC 
observations with IXAE on 1999, July 3 $-$ 13 and 1999, October 11 
$-$ 14 and the RXTE/ASM data since the beginning of the observation. 
From the calibration of the PPCs using Crab, it is found that the 
spectral data from PPC$-$3 is more reliable. Hence to obtain the 
hardness ratio (ratio of the count rate in 6 $-$ 18 keV energy range 
to the count rate in 2 $-$ 6 keV energy range), we have used the data 
from PPC$-$3. To study the spectral behavior of the source at different 
orbital phases, we have created the time averaged profile by folding the 
light curves and the hardness ratio data with a period of 17253 seconds. 
We have plotted, in Figure 1, the orbital modulated light curves in 2 $-$ 
18 keV and 6 $-$ 18 keV energy ranges and the hardness ratios for the 
1999, July and October observations of the source. 

The folded RXTE/ASM light curves in 3$-$5 keV and 5$-$12.1 keV energy bands 
plotted on a normalized scale with the template data and the corresponding 
hardness ratio for two different epochs (low and high states) are shown in 
Figure 3. From the PPC observations (Figure 1) and the RXTE/ASM observations 
(Figure 4), it is observed that, a strong binary phase dependence  of the 
spectrum is noticed.

\section{Discussion}

The nature of the compact object in the binary system of Cygnus~X-3 is not
yet clearly understood. There are various arguments in the literature 
supporting the compact object as a massive black hole whereas some authors 
believe the primary companion as a neutron star. Schmutz et al. (1996) 
suggested that the observed time variations in the profile of infrared 
emission lines from the binary system are due to the orbital motion of 
the companion which is a Wolf-Rayet star and derived  mass function for 
Cygnus~X-3 as 2.3 M$_{\odot}$. They obtained a range of 7$-$40 M$_{\odot}$ 
as the mass of the compact object with a most likely value of 17 M$_{\odot}$ 
and suggested that the binary system  Cygnus~X-3 is composed of a 
black hole (BH) and a Wolf-Rayet (WR) system. 
Several models have also been proposed to explain the lengthening of the 
orbital period (Bonnet-Bidaud \& Chardin, 1988). Conservative mass transfer 
from the companion star can not explain the observations. This is due to 
the fact that mass accretion rate needed to produce the amount of energy 
emitted in X-rays and the observed orbital period change rate requires 
a very low mass  companion $\sim$ 0.02 M$_{\odot}$, which will under-fill 
its Roche lobe to continue mass transfer. A low mass companion star with 
a moderate non-conservative mass loss rate, however, can explain the 
period change rate, if the mass loss is due to the X-ray heating of the 
atmosphere. On the other hand, if the companion star is massive, the orbit 
may widen only if there is significant mass loss from the system in the 
form of strong wind and the out-flowing mass carries specific angular 
momentum of the primary star. The emission line rich X-ray spectrum of
Cygnus~X$-$3 detected with Chandra (Paerels et al. 2000) indicates the
presence of strongly photo-ionized stellar wind reminiscent of a high mass
companion.

\section{Conclusion }
\begin{itemize}
\item We have measured 34 new minima and extended the measurement base to 
30 years.

\item The parabolic model provides the best fit to the new data with a modified
value of $\dot{P}$.

\item Sinusoidal model and the corresponding apsidal motion scenario is ruled out.

\item Mass loss from a massive companion is most favoured in Cygnus~X$-$3.
\end{itemize}

\section*{Acknowledgments  }
We thank the referees for their useful comments and suggestions that improved
the contents of the paper. We acknowledge the contributions of the scientific 
and technical staff of TIFR, ISAC and ISTRAC for the successful fabrication, 
launch and operation of the IXAE. We thank Dr. S. Seetha and Dr. K. 
Kasturirangan for their contribution to the IXAE. It is a pleasure to 
acknowledge the support of Shri K. Thyagarajan, Project Director IRS$-$P3 
satellite, Shri Tyagi, project manager of IRS-P3 and Shri J. D. Rao and 
his colleagues at ISTRAC. NSS thanks DST, Govt. of India, for providing 
the financial assistance. Work of SN is partially supported by the Kanwal 
Rekhi Scholarship of the TIFR Endowment Fund.

\end{document}